\documentclass[aps,preprint,groupedaddress]{revtex4}

\usepackage{graphicx}
\usepackage{graphics}

\begin{document}

\title{Analysis of realistic and empirical multipole interactions for shell-model calculations}
\author{K.~Sieja}
\address{Institute Pluridisciplinaire Hubert Curien, 23 rue du Loess, 
Strasbourg, France}
\date{today}

\begin{abstract}
We perform an analysis of realistic nucleon-nucleon 
interactions, as well as of empirically-corrected interactions,
fitted to reproduce in detail the spectroscopic data in $p$ and $sd$ shells. 
We focus on the multipole part of the interactions, contrary to 
previous studies of that type which focused on monopole parts. 
We analyze how the different parts of the NN force
(tensor, spin-orbit and central) contribute to the leading multipoles of the
interaction and how do they renormalize in medium. We show that the 
leading coherent terms of the realistic and effective interactions are 
dominated by the central component as expected from their schematic operator
structure. Convergence and cutoff dependence of multipole Hamiltonians is also discussed. 
   
\end{abstract}
%\pacs{21.60.Cs, 27.30.+t}

\maketitle
\section{Introduction}
One of the major problems in nuclear physics is the derivation of the 
realistic Hamiltonian in a given model space, that would be at the same time
consistent with the observed phase shifts and able to reproduce saturation properties 
of nuclei and their spectra. As these requirements have never been met in the calculations
with the realistic NN forces, the development of the NNN potentials on one side, 
and the phenomenological modeling of the nuclear Hamiltonians on the other, 
bring us closer and closer to a successful description of the atomic nucleus.
In the recent years a lot of attention has been paid to the monopole part of the effective interaction.
The first attempts have been undertaken to include some of the three-body effects 
into the calculation of 2-body monopoles in $sd$ \citep{Holt1} and $pf$ \citep{Holt2} shells,
showing that the effective 3-body $T=1$ interactions may be partially responsible for the observed spin-orbit 
shell closures. Otherwise, a lot of effort has been dedicated to the empirical modeling of the general monopole behavior,
which could be applied for studies of 
the shell evolution far from stability. An early trial in Ref. \citep{Otsuka2001} 
consisted in explaining the shell closure $N=16$ with a central spin-isospin exchange term.
Further a dominant role of the tensor force has been noticed
\citep{Otsuka2006}, however the tensor scenario holding in light nuclei seemed to contradict the shell 
evolution and stability of heavier systems \citep{Nowacki:2007app, Sieja-zr}. 
A quantitative analysis 
of the role played by different components of the NN interaction in the $sd-pf$  shell was done \cite{Nadya2010}
using the spin-tensor decomposition of the 2-body effective interactions. 
It has been shown that the evolution of the $N=16,20$ and $N=28$ shell gaps is in fact due to the
combined effects of the central and tensor forces. In Ref. \cite{Otsuka2010} a very same conclusion was
achieved in yet another empirical attempt of providing a universal monopole interaction.
It was also noted in Ref. \cite{Otsuka2011} that the tensor component of the monopole field remains nearly intact
in the perturbative treatment, i.e. the monopole tensor content of the realistic NN force
remains the same in the effective interactions. It is of interest to notice that
the tensor contribution to the monopole seems to survive intact even in the empirical fits, as presented e.g.
in \cite{F-Trento}.
This feature of the tensor force, termed {\it renormalization persistency} in Ref. \cite{Otsuka2011}, was studied 
in detail only for the monopole matrix elements of the effective Hamiltonian.
It has been however noted  that considerable differences appear in the non-diagonal elements of 
the bare and renormalized tensor forces, whose consequences were not studied in the shell model. 

While the saturation properties of nuclei are determined by the monopole part of the realistic forces,
the detailed spectroscopy is dependent not only on the monopole field propagation but
as well on the multipole part of the interaction, the pairing force determining 
the $0^+-2^+$ energy splittings in a core$\pm$2 particles system being the most prominent example.
The addition of the core polarization corrections to the realistic two-body matrix 
elements by Kuo and Brown \cite{Kuo68} yielded a first reasonable agreement of spectra of even-even systems with data.
Though the perturbative treatment of the in-medium effects have been further developed \cite{Kuo90,Jensen94,Jensen95},
it is commonly recognized that an accurate spectroscopic description of nuclei 
still requires a simultaneous readjusting of the monopole and multipole interactions. 
To which extent the deficiencies in the multipole part come from the lack of many-body effects in the initial potential
or/and from the shortcomings of the perturbative treatment, remains also an open question.

The multipole part of the Hamiltonian was less extensively studied than the monopole one, with the ground-breaking
exception of Refs. \cite{Zuker94, Dufour:1995em}, where the leading multipoles in the effective interactions
were classified and the "universality" of these multipoles in different
realistic interactions evidenced, i.e. the fact that the multipole content of all the realistic interactions is fairly similar.
The problem of the $T=1$ multipole component of the effective interactions far from stability
was addressed e.g. in \cite{corep}, showing that a proper renormalization is crucial
for achieving a correct magnitude of multipole components in $p-sd$ and $sd-pf$ interactions.

In this work, we concentrate as well on the multipole part of the effective interactions,
this time we analyse both $T=0$ and $T=1$ components in one major harmonic oscillator shell
($p$ and $sd$ shells). For the first time,
we perform the spin-tensor analysis of the multipole terms. 
The effective interactions obtained from the CD-Bonn potential in the many-body perturbation theory are further compared
to the phenomenological ones, based on the very same matrix elements and 
fitted to experimental data using the linear combination method, as practiced e.g. in \cite{USDB, Honma2009, Gniady}. 

We start the discussion with a short reminder of the multipole-monopole decomposition of
the interaction as well as the spin-tensor decomposition in Sec. \ref{sec-decomp}.
In Section \ref{sec-fit} we give some details of the derivations and fits of the effective interactions.
Finally, we  perform the spin-tensor analysis of the realistic 
and readjusted interactions in Sec. \ref{sec-res} and we present the major
results of this work. Conclusions are collected in Sec. \ref{sec-conc}.

\section{Spin-tensor and monopole-multipole decomposition of the effective Hamiltonian\label{sec-decomp}}
The spin-tensor decomposition is a procedure allowing
to separate the central, spin-orbit and tensor parts of the effective interaction. The monopole
and multipole parts of each component can be then studied unambigously.
The most general 2-body effective interaction can be written as
\begin{equation}
V(1,2)=\sum_{k=0,1,2} V^{(k)},
\end{equation}
where $k=0$ stands for the central, $k=1$ for the spin-orbit and $k=2$ for the tensor part
of the interaction. 
To extract each component from the $J$-coupled two-body matrix elements (TBME), commonly used in SM
studies,  one needs to pass from the $J$ to the $LS$ coupling and then employ the spin-tensor decomposition
to the $LS$-coupled matrix element. We thus have:

\begin{eqnarray}
\langle ab; JT |V_k| cd; JT\rangle=
\frac{(2k+1)(-)^J}{\sqrt{(1+\delta_{ab})(1+\delta_{cd})}}
\sqrt{(2j_a+1)(2j_b+1)(2j_c+1)(2j_d+1)} \nonumber \\
\times \left[\sum_{LS,L'S'}(2L+1)(2S+1)(2L'+1)(2S'+1) 
\left\{\begin{array}{ccc} l & 1/2 &j_a \\
                         l & 1/2 &j_b \\
			 L & S & J
			 \end{array}\right\} 
\left\{\begin{array}{ccc} l & 1/2 &j_c \\
                         l & 1/2&j_d \\
			 L' & S' & J
			 \end{array}\right\}\right.   \nonumber\\			 
\left\{\begin{array}{ccc} L & S & J \\
                         S' & L' & k 
			 \end {array}\right\} \times \left\{\sum_{J'}(2J'+1)(-^{J'})\left\{\begin{array}{ccc} L & S & J' \\
                         S' & L' & k 
			 \end {array}\right\} \right .  \nonumber \\
\times \left( \sum_{j_a' j_b' j_c' j_d'}	\sqrt{(2j_a'+1)(2j_b'+1)(2j_c'+1)(2j_d'+1)}\left\{\begin{array}{ccc} l & 1/2 &j_a' \\
                         l & 1/2 &j_b' \\
			 L & S & J'
			 \end{array}\right\} 
\left\{\begin{array}{ccc} l & 1/2 &j_c' \\
                         l & 1/2&j_d' \\
			 L' & S' & J'
			 \end{array}\right\}	\right .\nonumber	\\		 	 
 \times \sqrt{(1+\delta_{a'b'})(1+\delta_{c'd'})}\langle a'b';J'T|V|c'd'; J'T\rangle \Bigg)\Bigg \} \Bigg],			 
\end{eqnarray}
where $\langle a'b';J'T|V|c'd'; J'T\rangle$ are the $J$-coupled TBME of the interaction to be decomposed. 

To asses the properties of the multipoles of any interaction, we need to decompose
the Hamiltonian into the multipole and monopole part.
The idea of the multipole-monopole separation is that the Hamiltonian can be 
written in terms of the density operators coupled to a good angular momentum
$\lambda$, i.e. $(a^\dagger a)^\lambda$ and that the monopole component
is given by $\lambda=0$ while the multipole is all the rest $\lambda\ne0$ \cite{Dufour:1995em}.
A proof of the separation property is given in Ref. \cite{Zuker94}. 
$H_{multipole}$ takes the following form in the particle-particle representation:
\begin{equation}
H_{multipole}=\sum\limits_{ik<jl\Gamma}
W_{ijkl}^\Gamma [(a_i^\dagger a_j^\dagger)^\Gamma (a_ka_l)^\Gamma]^0,
\end{equation}
where $\Gamma=JT$, and in the particle-hole:

\begin{eqnarray}
H_{multipole}=\sum_{ik<jl \Gamma}\sqrt{(2\gamma+1)}{\sqrt{(1+\delta_{ij})
(1+\delta_{kl})}\over4}\omega_{ikjl}^\gamma [(a_i^\dagger\tilde a_k)^\gamma
(a_j^\dagger \tilde a_l)^\gamma]^0,
\end{eqnarray}
where $\gamma=\lambda\tau$. We have replaced here $V$ matrix elements by $W$, to underline that
they are monopole-free. The $W$ and $\omega$ matrix elements are related to each other by a Racah
transformation:
\begin{eqnarray}
\omega_{ikjl}^\gamma =\sum_\Gamma(-)^{j+k-\gamma-\Gamma}
\left\{\begin{array}{ccc}
   i&j&\Gamma\\  l&k&\gamma
   \end{array}\right\}
W_{ijkl}^\Gamma(2J+1)(2T+1),\label{3a}
\end{eqnarray}
\begin{eqnarray}
W_{ijkl}^\Gamma =\sum_\gamma(-)^{j+k-\gamma-\Gamma}
\left\{\begin{array}{ccc}
   i&j&\Gamma\\  l&k&\gamma
   \end{array}\right\}
\omega_{ikjl}^\gamma(2\lambda+1)(2\tau+1).\label{3b}
\end{eqnarray}
The knowledge of the multipole Hamiltonian is connected to a large number of the matrix elements
among which some may be more important than others. 
The diagonalization of the multipole Hamiltonian in the particle-particle
and particle-hole representations allow to pin down the coherent terms and to separate 
the collective part (C) from the random one (R), both contributing to the multipole Hamiltonian: $H_{multipole}=H_C+H_R$.
The analysis of the strongest contributions appearing in the diagonal form of $H_{multipole}$ was performed in Ref. \cite{Dufour:1995em}.
In this work we focus on the two dominant $JT=10,01$ terms, corresponding to the pairing operators $P^{10},P^{01}$, and four  
representative $\lambda\tau$: 10 ($\sigma$), $11$ ($\sigma\tau$), $20$ ($r^2 Y_2$) and 30 $(r^3 Y_3)$.
The proposed operator structure of these multipoles
assumes that they are central and the overlaps of the schematic operators with their realistic multipole 
counterparts were shown to be quite close to 1 (see Table II from Ref. \cite{Dufour:1995em}). 
Thus the magnitude of the tensor and spin-orbit components in the spin-tensor decomposition of the multipole Hamiltonian
will provide yet another measure of the possible deviation of the collective terms from their analytical forms.

\section{Fit of the interaction\label{sec-fit}}
The starting point of the fit procedure is an effective Hamiltonian  
obtained from the CD-Bonn potential \cite{CD-Bonn}, renormalized via $V_{lowk}$ approach with a sharp cutoff \cite{vlowk}
of 2fm$^{-1}$. The $V_{lowk}$ is solved by the similarity transformation in the oscillator space of 10 major shells
with $\hbar\omega=18$ for $p$ nuclei and  $\hbar\omega=15$
for $sd$ nuclei. Effective Hamiltonians were further evaluated
through the $\hat Q$-box method with folded diagrams as described in Refs. \cite{Kuo90, Jensen94, Jensen95}. 
All diagrams through third order
have been included, allowing $18\hbar\omega$ excitations for intermediate states. 
For the calculations of the $V_{lowk}$ potential and effective Hamiltonians
the Oslo codes \cite{MHJwebsite} were employed.
For the purpose of the fit with the linear combination method we selected
41 energy levels in 14 nuclei in the $p$ shell and 300 levels in 59 nuclei in the $sd$ shell.
For such data sets, the rms deviations obtained with effective interactions calculated with the $\hat Q$-box
method are 2.74MeV and 2.29MeV, respectively. It is not encouraging to notice, that on average similar results 
are obtained using bare $V_{lowk}$ interactions (rms =2.09MeV in $p$ and 2.18MeV in $sd$ shells). However, as will be discussed later,
the particular properties of the interaction (e.g. pairing) are much better given by the effective $\hat Q$-box
interaction.

In the first step of the fit we adjusted only the single particle energies and monopoles. 
The fit was achieved using 4 linear combinations for the $p$-shell
and 12 linear combinations for the $sd$ shell. The rms deviations obtained this way are 1.74MeV and 0.83MeV for the $p$ and $sd$
shells, respectively. The improvement with respect to the $\hat Q$-box interactions is considerable, still the precision
of the interaction is far from being satisfactory.
In the second step we have used the monopole-tuned interactions as a starting point
of a full-fledged fit of both monopole and multipole components. In this case, we have obtained
the best agreement with data using 10 linear combinations for the $p$ shell and 40 linear combinations 
for the $sd$ shell. The simultaneous fit of the monopole and multipole components allow to 
improve greatly the agreement with data, leading to rms deviations of 0.61MeV and 0.22MeV accordingly.
One should however state, that it was not our purpose to obtain the best up-to-date fit of the effective
interactions, but rather to provide consistent phenomenological Hamiltonians for the sake of further analysis.

\section{Results\label{sec-res}}
\subsection{Spin-tensor analysis}
\begin{figure}
\begin{center}
\includegraphics[scale=0.5]{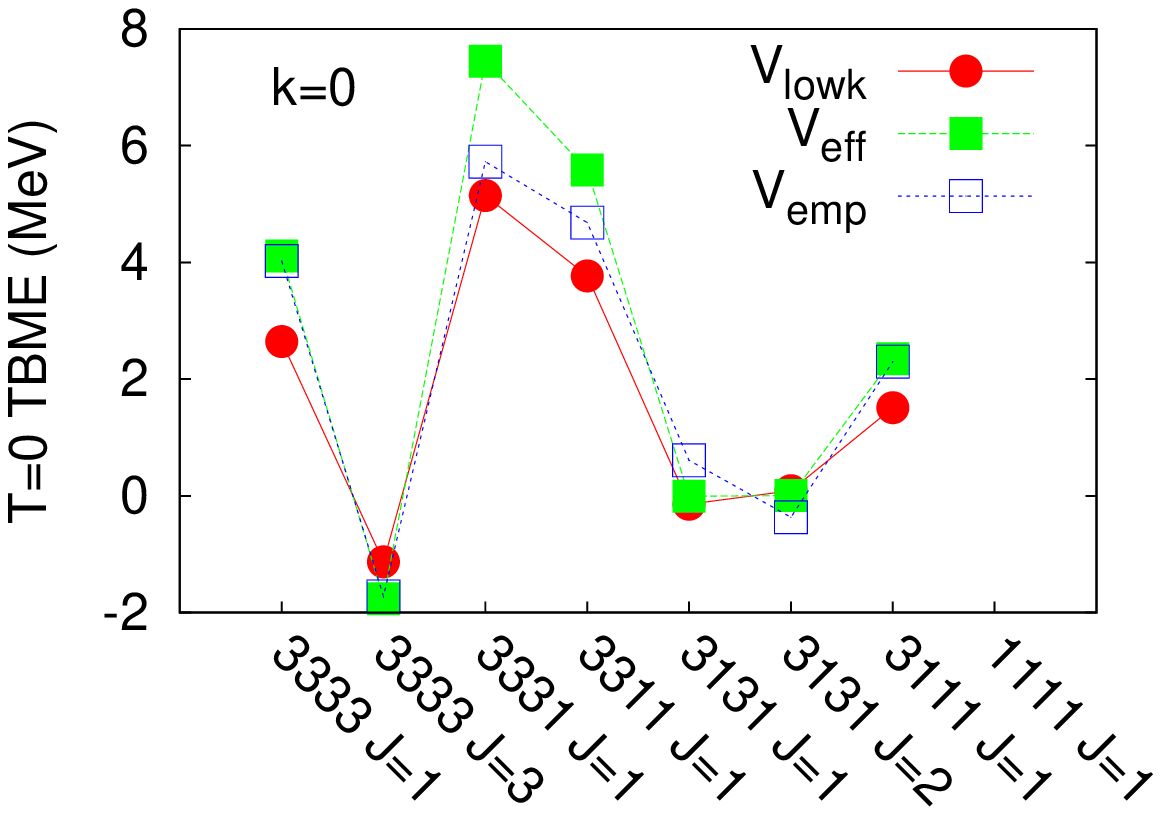}\\
\includegraphics[scale=0.5]{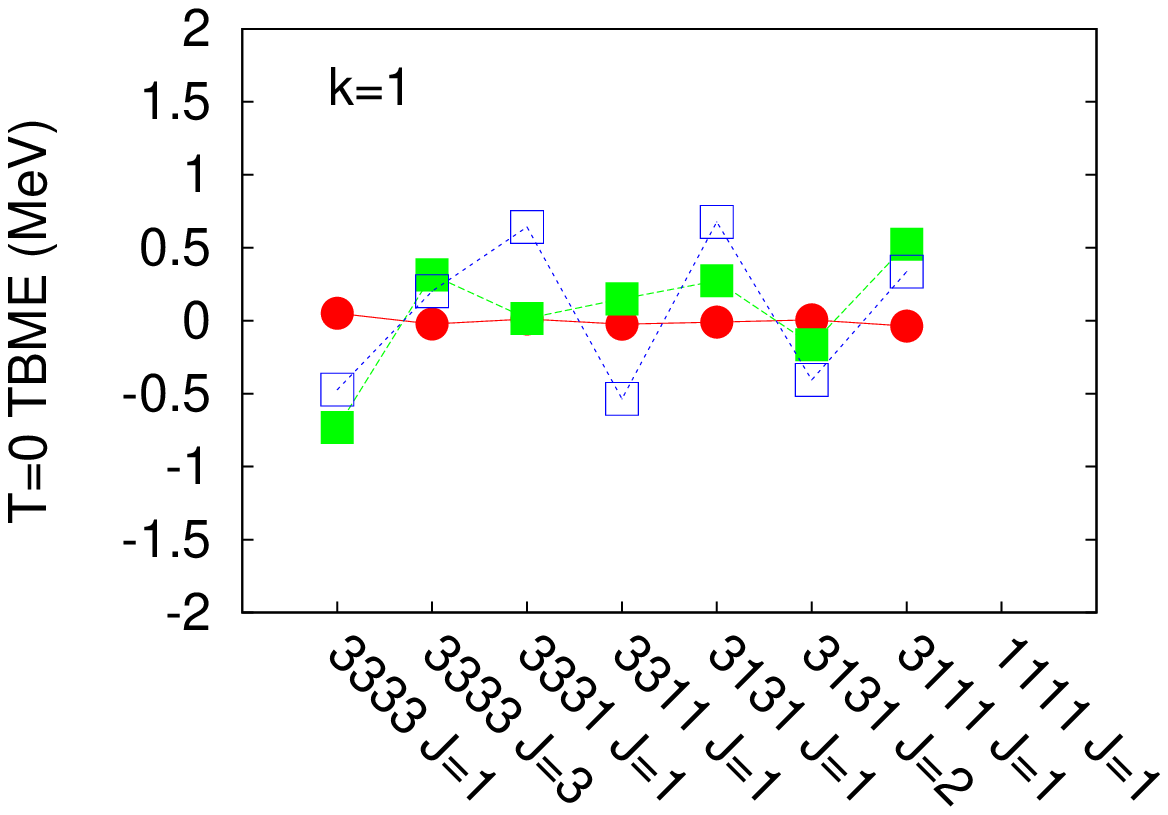}\\
\includegraphics[scale=0.5]{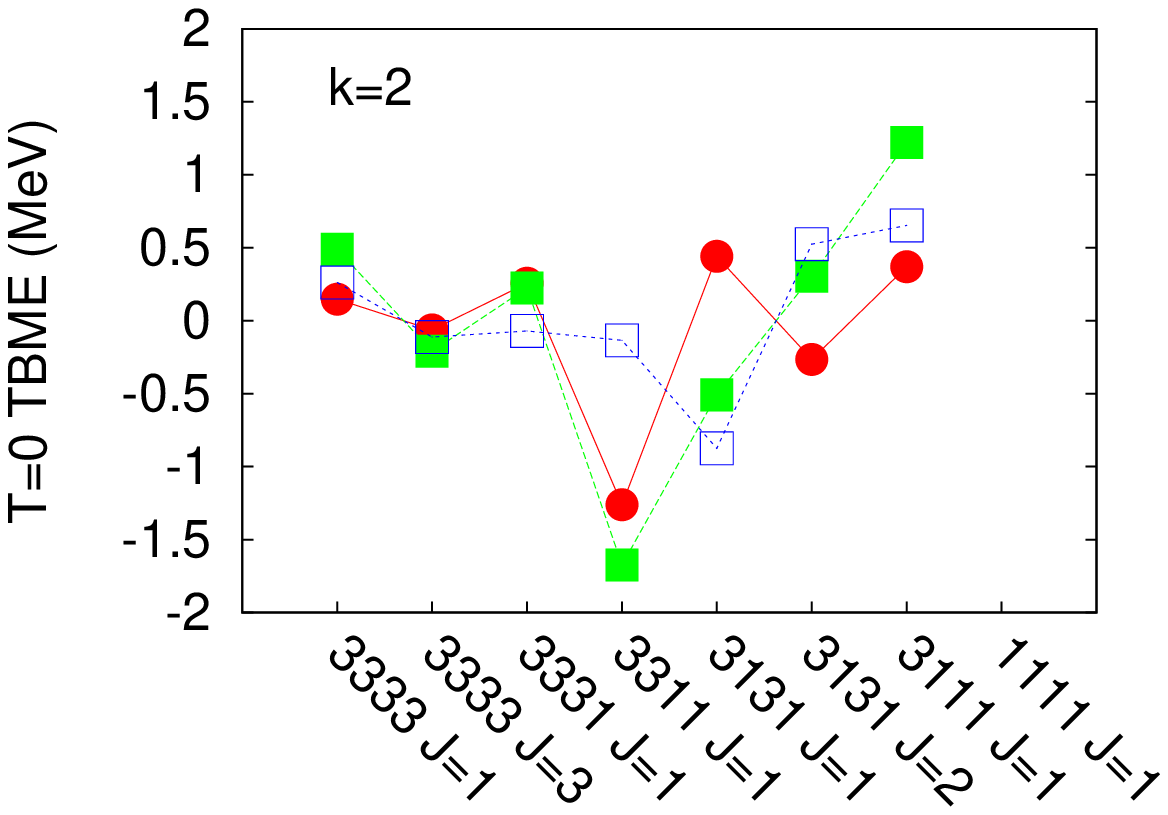}
\end{center}
\caption{Matrix elements of the $p$ shell monopole-free interaction. 
$V_{lowk}$ potential (first order interaction) 
is compared to the effective Hamiltonian ($V_{eff}$) calculated in the 3rd order $\hat Q$-box method
including folded diagrams and to the empirical fit ($V_{emp}$). Central ($k=0$), spin-orbit ($k=1$) and tensor ($k=2$)
components are disentangled. $3$ stands here for the $p_{3/2}$ and $1$ for the $p_{1/2}$ orbital.}\label{fig-meten}
\end{figure}

We start the discussion with the TBME of the isoscalar interaction in the $p$ shell.
In Fig. \ref{fig-meten} we show the matrix elements of the $p$ shell interactions
in which the monopole component was removed. We compare the first order interaction
($V_{lowk}$) to the 3rd order $\hat Q$-box interaction with folded diagrams ($V_{eff}$)
and the empirical fit ($V_{emp}$).
First of all, let us notice that the monopole-free matrix elements have the largest central
part and smaller, but non-negligible spin-orbit and tensor components.
All the three components are modified due to the in-medium corrections, and in particular, 
the tensor component of the multipole part of the $T=0$ interaction
is equally renormalized.
This is not the case of the monopole interaction, in which the tensor part remains
nearly the same, as was shown e.g. in Ref. \cite{Otsuka2010}.
Thus the renormalization persistency does not hold for the monopole-free TBME
of the interaction.
It is also worth noticing, that the empirical TBME follow
the global trends of monopole-free TBME of the $V_{lowk}$ and the fit leads to moderate changes in their amplitude,
a condition necessary for a "realistic-compatible" interaction. 

The multipole Hamiltonian, whose TBME are plotted in Fig. \ref{fig-meten},
can be however split into random and collective parts. For further discussion 
we will pass to the diagonal representation as discussed in Sec. \ref{sec-decomp}
and analyse rather the coherent multipole terms than the monopole-free 2-body matrix elements one by one.

In Tables \ref{tab-p},\ref{tab-sd} we show the eigenvalues of multipole operators
in the first ($V_{lowk}$) and third order $\hat Q$-box with folded diagrams ($V_{eff}$) interactions compared to the empirical fits ($V_{emp}$) 
in $p$ and $sd$ shells.
We diagonalize separately central, spin-orbit and tensor components in each case. 

\begin{table}
\begin{center}
\caption{Eigenvalues of the leading multipoles in the $p$-shell. Central, spin-orbit and tensor components
of the CD-Bonn potential smoothed with the $V_{lowk}$ procedure,
3rd order interaction with folded diagrams ($V_{eff}$) and the empirical fit ($V_{emp}$) 
are distinguished.\label{tab-p}}
\begin{tabular}{cccccc}
\hline
\hline
Multipole & Interaction & Central & Spin-orbit & Tensor & Total \\
\hline
$JT=10$           & $V_{lowk}$  & -4.46   &   -0.04   &  -1.30     &   -4.00\\
                  & $V_{eff}$  & -6.23   &   -0.77   &  -2.17     &   -6.28\\
                  & $V_{emp}$  & -4.11   &   -1.11   &  -1.22     &   -4.42            \\
\hline
$JT=01$         & $V_{lowk}$   & -4.88   &   -0.30   &  -0.85     &   -5.72\\  
                & $V_{eff}$  & -6.77   &   -0.20   &  -0.58     &   -7.16\\
                & $V_{emp}$  & -6.85   &   -0.16   &  -0.72     &   -6.83 \\
\hline
$\lambda\tau=10$ & $V_{lowk}$ &  -1.71& -0.44 &  -0.87 & -1.90\\
                 & $V_{eff}$&    -2.22& -0.29 &  -0.66 & -2.59\\  
                 & $V_{emp}$&    -1.67& -0.48 &  -0.48 & -2.00 \\
\hline
$\lambda\tau=20$ & $V_{lowk}$ & -3.06  &  -0.26   &  -0.34 &  -3.05\\
                 & $V_{eff}$&   -4.31  &  -0.16   &  -0.07 &  -4.26\\
                 & $V_{emp}$&    -3.56 &  -0.22   &  -0.30 &  -3.59\\
\hline 
$\lambda\tau=11$ & $V_{lowk}$ & 2.43   &  0.15    &  0.19  &  2.35\\
                 & $V_{eff}$& 3.51     &  0.21    &  0.17  &  3.49\\
                 & $V_{emp}$& 2.93     & 0.09     &  0.24  &  2.96\\
\hline        
$\lambda\tau=30$ & $V_{lowk}$ & -0.67  &  0.09   &  0.06 &  -0.52\\
                 & $V_{eff}$& -0.98    &  0.20   &  0.00  & -0.78 \\
                 & $V_{emp}$& -1.02    &  0.33   &  0.03 &  -0.65\\
\hline
\hline
\end{tabular}
\end{center}
\end{table}

\begin{table}
\begin{center}
\caption{The same as in table \ref{tab-p} but for the $sd$ shell.\label{tab-sd}}
\begin{tabular}{cccccc}
\hline
\hline
Multipole & Interaction & Central & Spin-orbit & Tensor & Total \\
\hline
$JT=10$        &  $V_{lowk}$ & -4.78  & -0.02   & -0.98  &    -4.93\\
          &  $V_{eff}$  & -7.24  &   -0.62   &    -2.02  &    -8.08\\
          &  $V_{emp}$  & -5.61  &   -0.61   &    -2.03  &    -6.58\\ 
\hline
$JT=01$ &  $V_{lowk}$   & -4.18  &   -0.02   &    -0.24  &   -4.41\\
        &  $V_{eff} $   & -6.37  &   -0.02   &    -0.15  &   -6.38\\
        &  $V_{emp}$    & -5.76  &   -0.13   &    -0.10  &   -5.80\\
\hline
$\lambda\tau=10$& $V_{lowk}$ & -1.38 & -0.32  &   -0.71  &    -1.53\\
                & $V_{eff}$  & -1.91 & -0.19  &   -0.77  &    -2.09\\     
                & $V_{emp}$  & -1.22 & -0.89  &   -0.43  &    -1.66\\
\hline
$\lambda\tau=20$& $V_{lowk}$ & -2.85 &    -0.31 &      -0.35 &   -2.86\\
                & $V_{eff}$  & -4.47 &    -0.69 &      -0.17 &   -4.54\\
                & $V_{emp}$  & -3.38 &    -0.52 &      -0.16 &   -3.44\\
\hline
$\lambda\tau=11$& $V_{lowk}$ & 2.77  &   0.10  &    0.24 & 2.75\\ 
                & $V_{eff}$  & 4.11  &   0.10  &    0.28 & 4.09\\
                & $V_{emp}$  & 3.81  &   0.28  &    0.31  & 3.85\\
\hline
$\lambda\tau=30$& $V_{lowk}$ & -0.75 & -0.23    & -0.48  &  -0.77 \\
                & $V_{eff}$  & -0.99 & -0.22    & -0.23  &  -1.10 \\
                & $V_{emp}$  & -0.73 & -0.23    & -0.34  &  -1.04 \\ 
\hline
\hline
\end{tabular}
\end{center}
\end{table}

Several points need to be noted from Tables \ref{tab-p}, \ref{tab-sd}:

(i) The central parts of the multipole interactions have the eigenvalues very close to
that of the total multipole Hamiltonian. A fact expected from \cite{Dufour:1995em}
and supporting the operator forms proposed for the leading multipole terms. 

(ii) It is interesting to notice that the eigenvalues of tensor and spin-orbit components
in $JT=10$ and $\lambda\tau=10$  are the largest and these are the cases for which the 
weakest overlaps with the schematic operators were found in Ref. \cite{Dufour:1995em}. 

(iii) In-medium corrections and empirical fits affect all the multipole components in all its 
spin-isospin structures, though the change 
is of course the most visible in the largest central part. 
Even if the renormalization persistency
of the tensor force does not hold for the multipole part, this is of minor importance for nuclear modeling.
The first order tensor contribution in the multipole Hamiltonian can be simply neglected.

(iv) In the majority of cases, the fit tends to reduce the multipole component
to a value somewhere in between the $V_{lowk}$ and $V_{eff}$. 
Roughly the eigenvalues of the empirical interaction can be obtained from $V_{lowk}$/$V_{eff}$ by multiplying/dividing
its TBME by a factor 1.2 in both shells. 

(v) We presented the results in $p$ and $sd$ shells only but given the properties of
the multipole Hamiltonian \cite{Dufour:1995em}, the analysis
in any other shell or in multi-shell valence spaces could yield
the same conclusions.

\subsection{Convergence of multipoles}

\begin{figure}
\begin{center}
\includegraphics[scale=0.35]{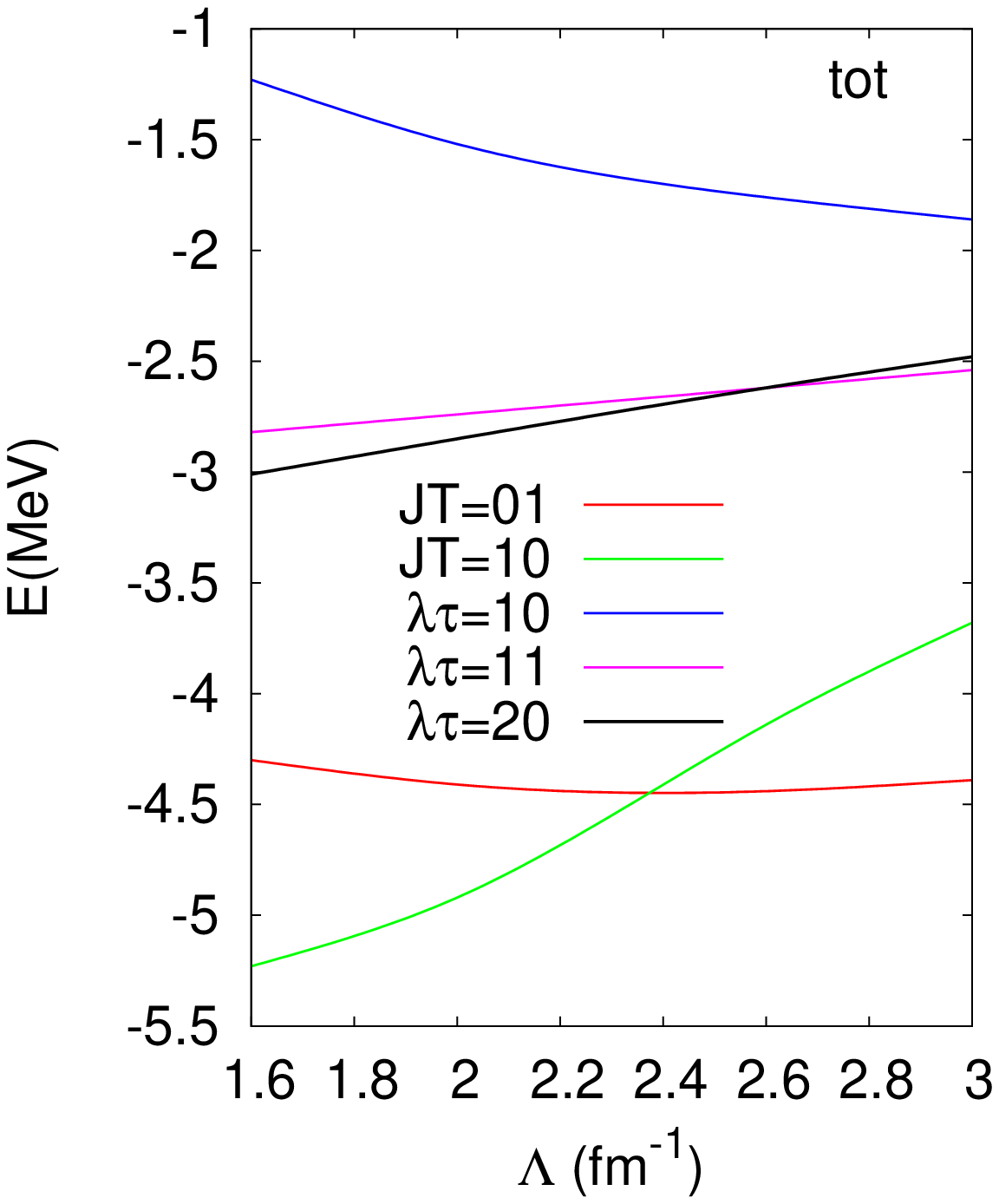}\includegraphics[scale=0.35]{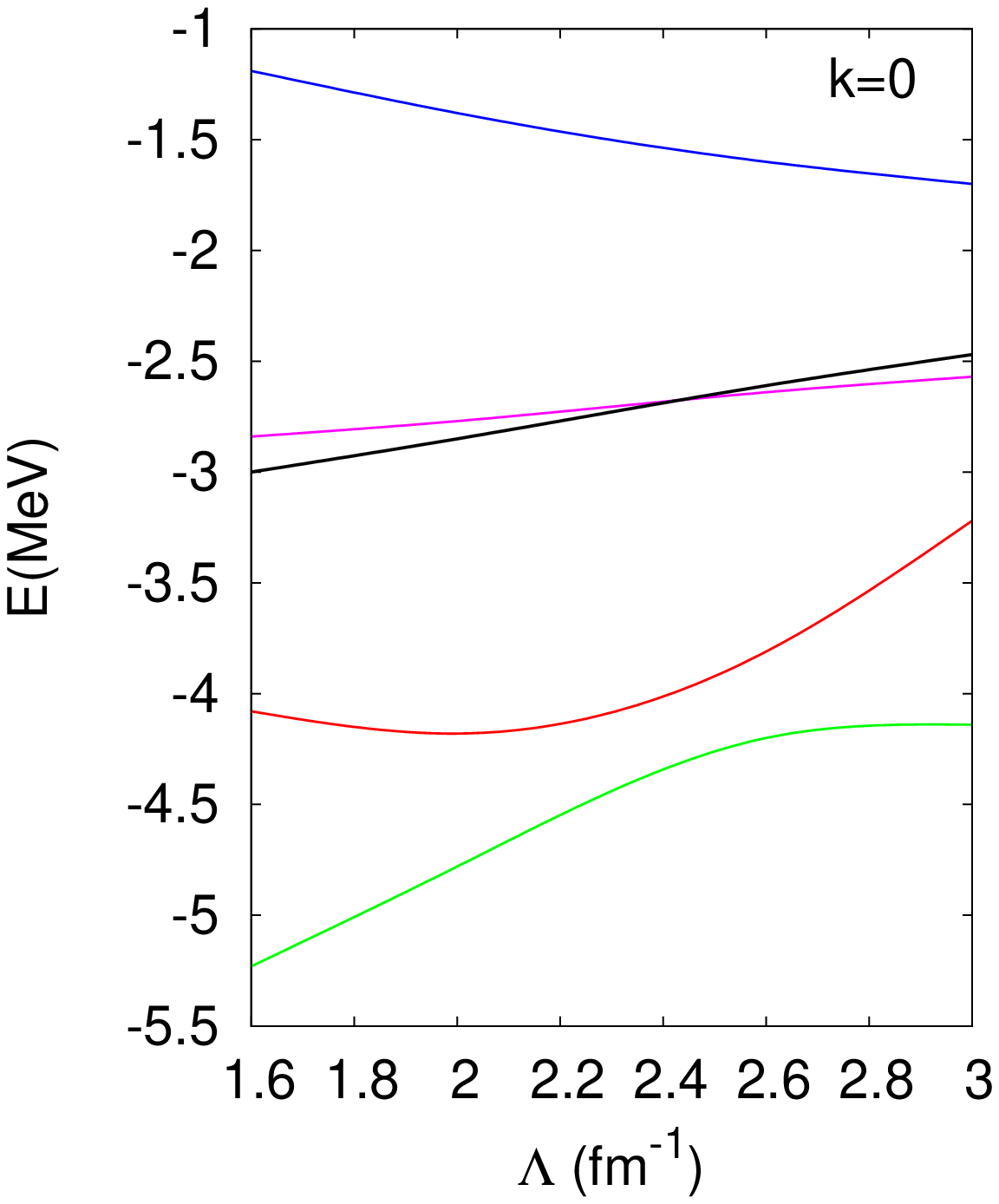}\\
\includegraphics[scale=0.35]{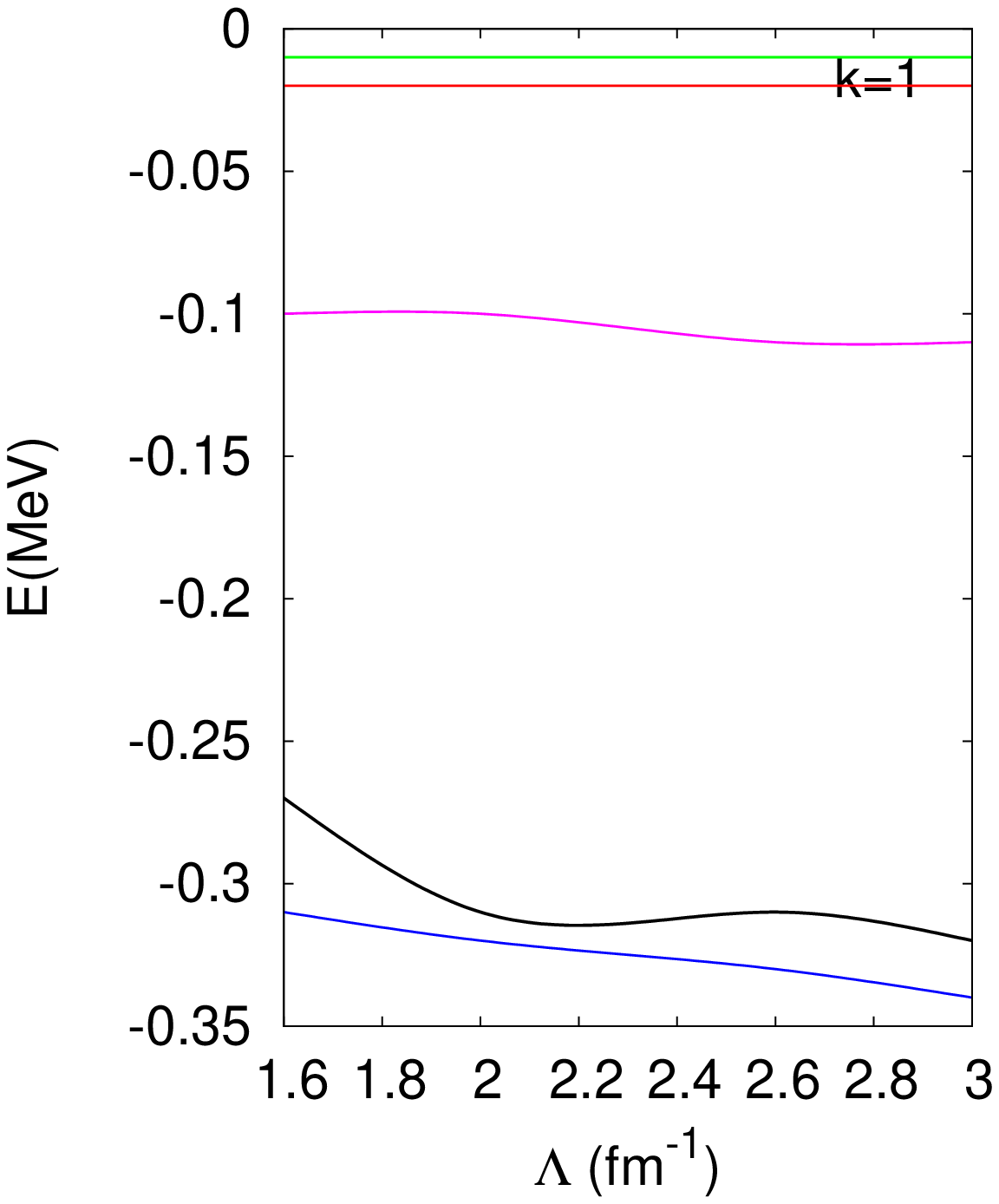}\includegraphics[scale=0.35]{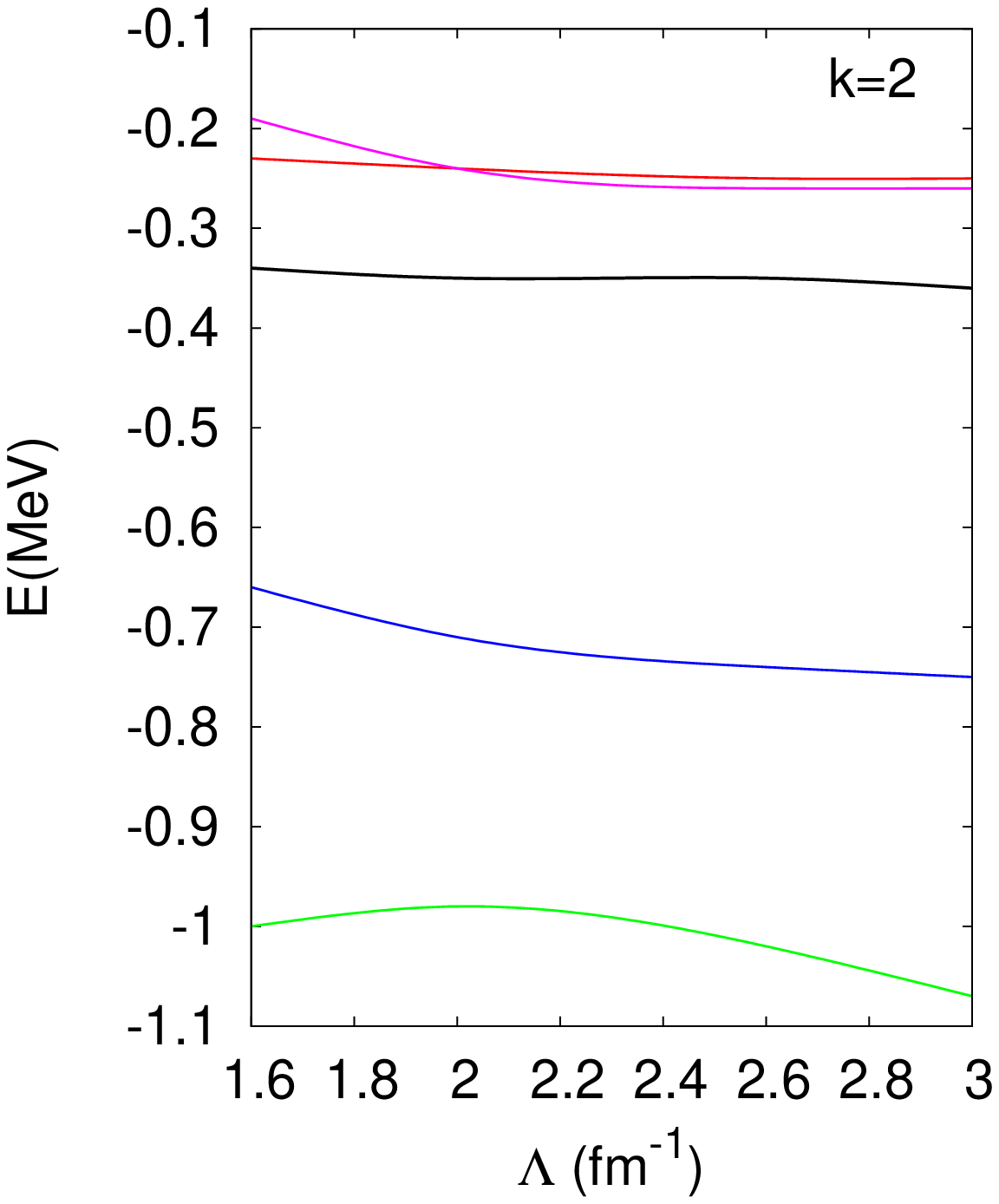}
\caption{Multipole eigenvalues in the $V_{lowk}$ potential and its central, vector and tensor
components, depending on the cutoff $\Lambda$
The value of $\lambda\tau=11$ is multiplied by -1. See text for details.\label{fig-vlowk}}
\end{center}
\end{figure}

So far, we did not discuss the cutoff and convergence properties of the multipole terms in different interactions.
Examining the cutoff dependence of the calculated observables may be a useful way to estimate the uncertainty of the results
due to the neglected many-body forces and the problems in the perturbative treatment. In Ref. \cite{corep} 
we noticed that the cutoff dependence of multipoles is rather weak in the range of $1.8-2.6$fm$^{-1}$, at least
for pairing and quadrupole in the $T=1$ channel. Here we repeated this analysis for all the multipoles
in the wide range of $\Lambda=1.6-2.6$fm$^{-1}$. The result is shown in Fig. \ref{fig-vlowk}
for the multipoles of the $V_{lowk}$ potential in the $sd$-shell, distinguishing as well cutoff variation of 
the central, spin-orbit and tensor components.
We have multiplied the value of the $\lambda\tau=11$
by -1 and skipped the $\lambda\tau=30$ for a better transparency in the Figure. 
The omitted octupole term shows a quite weak cutoff dependence changing its value on about $10\%$
within the examined cutoff range. 
As can be seen from the Figure, the $JT=10$ shows the strongest cutoff dependence. 
The strong cutoff dependence of 
the $T=0$ TBME was attributed to the contribution of the short-range tensor force.
One can also note, what was already apparent from the Tables \ref{tab-p}, \ref{tab-sd}, that the 
multipoles of the $V_{lowk}$ are basically given by its central component, which has the strongest
variation with $\Lambda$. The negligible vector and a bit larger tensor component are 
nearly cutoff independent (note however different energy scales in the Figure). 

\begin{figure}
\begin{center}
\includegraphics[scale=0.4]{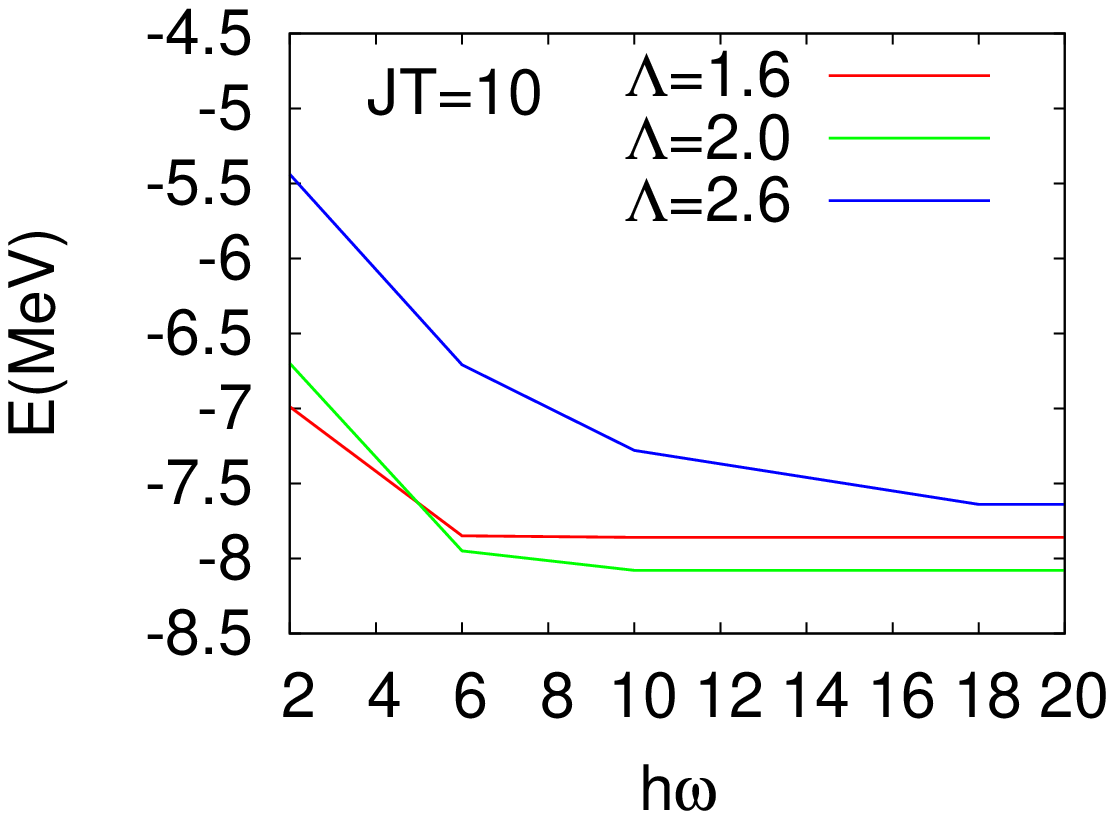}\includegraphics[scale=0.4]{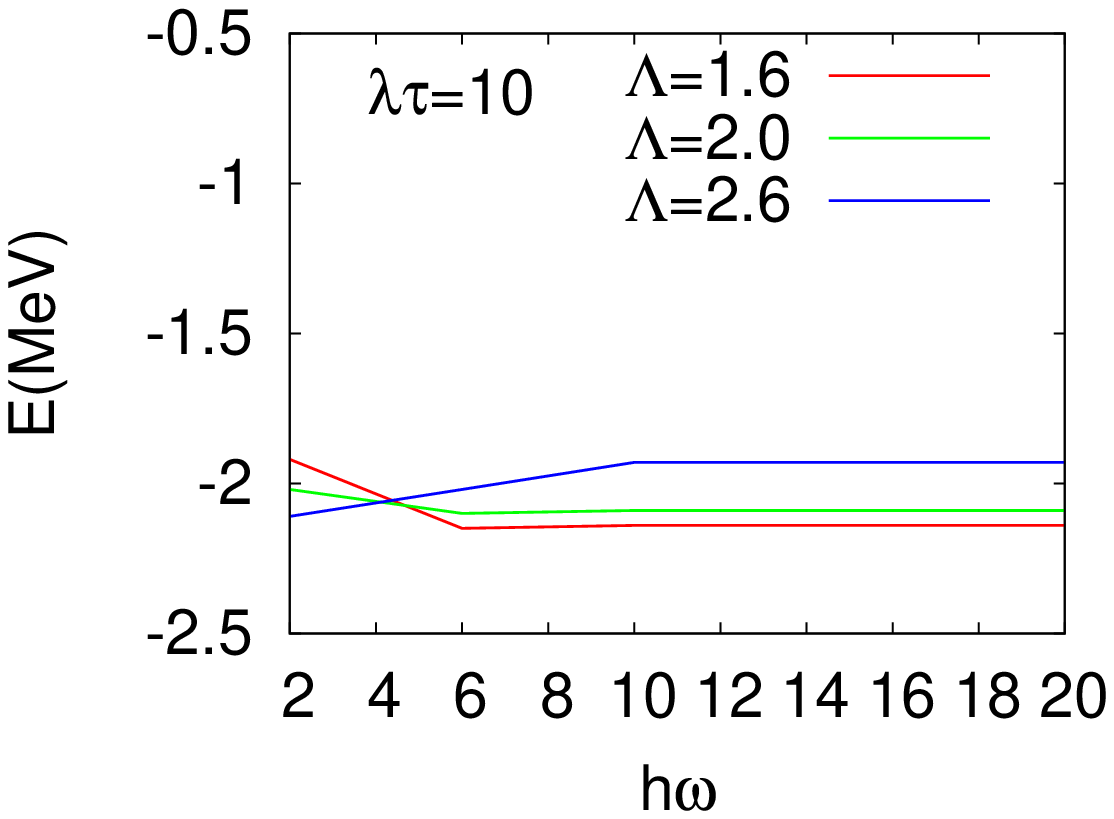}\\
\includegraphics[scale=0.4]{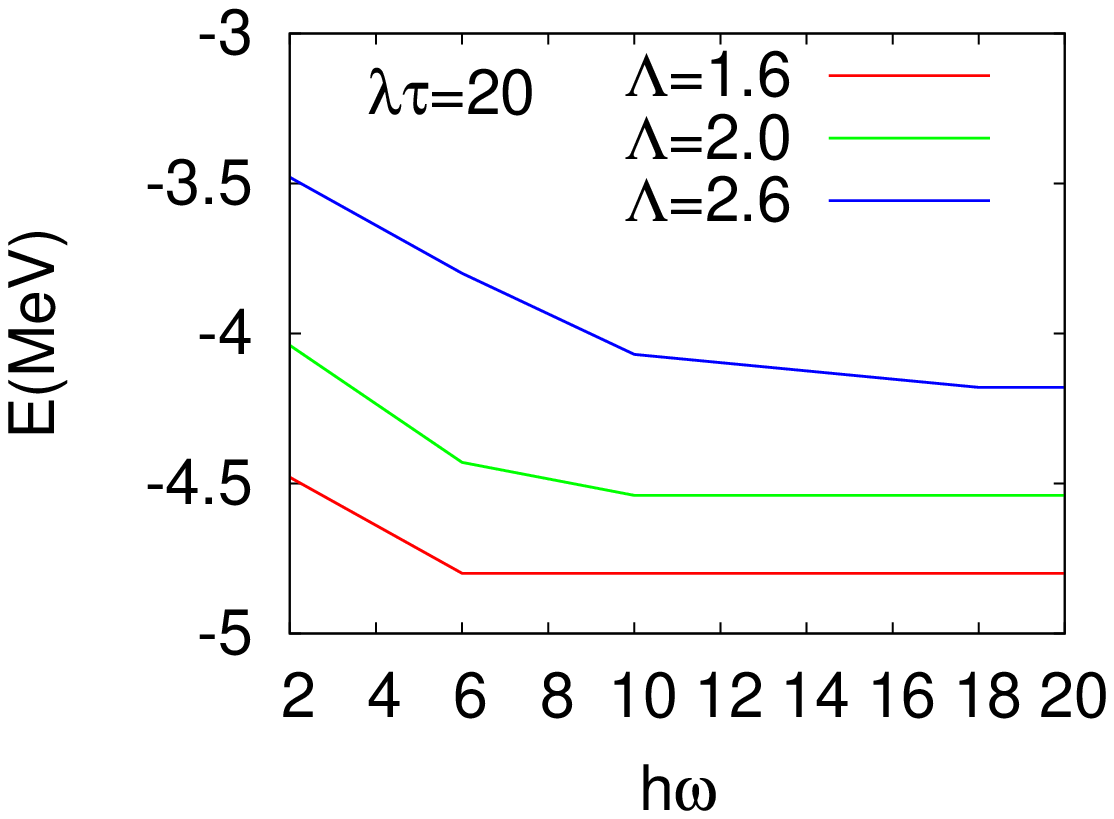}\includegraphics[scale=0.4]{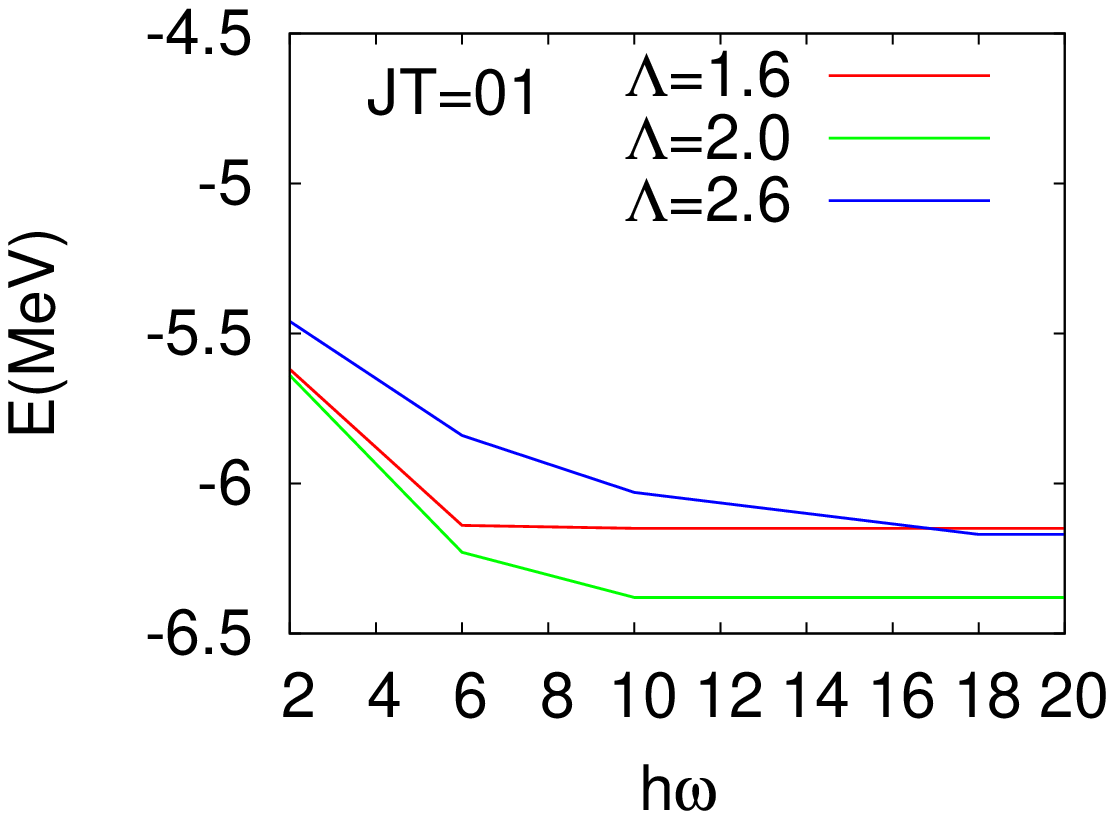}\\
\includegraphics[scale=0.4]{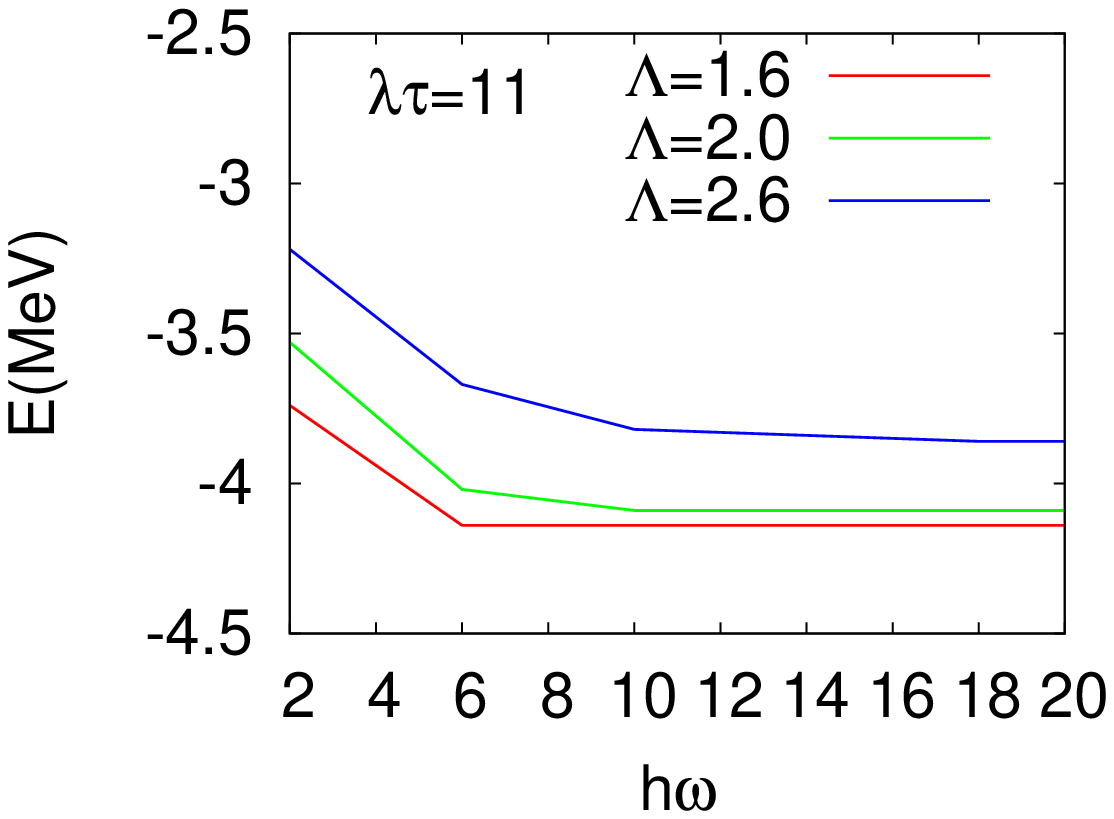}\includegraphics[scale=0.4]{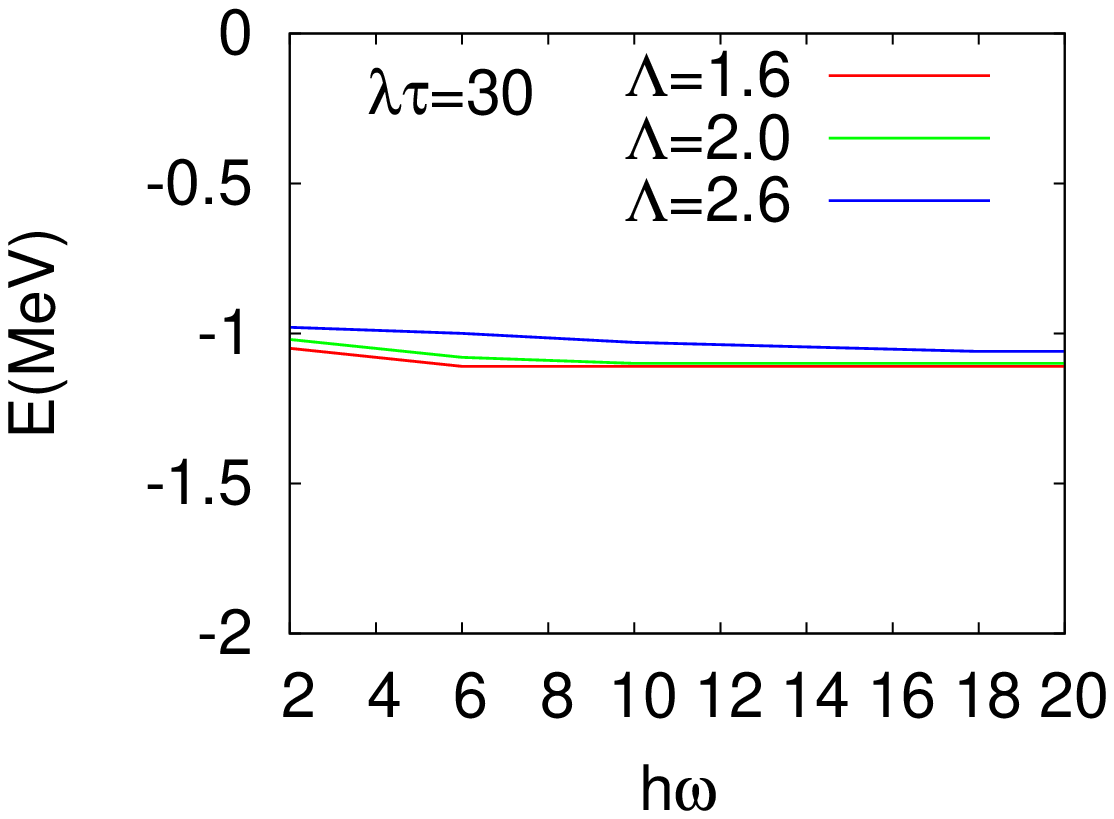}
\caption{Multipole Hamiltonian eigenvalues plotted vs $\hbar\omega$ excitations of 2-body diagrams included in the evaluation of $\hat Q$-box
for different cutoffs of the starting $V_{lowk}$ potential. $\lambda\tau=11$ value is multiplied by -1. Note different energy range
in the case of $JT=10$. \label{fig-cut}}
\end{center}
\end{figure}

In Fig. \ref{fig-cut} we show the convergence of multipoles of an effective Hamiltonian in the $sd$-shell, this time
eigenvalues of multipoles calculated to 3rd order in $\hat Q$-box are plotted against the number of $\hbar\omega$
excitations permitted for intermediate diagrams, obtained with the $V_{lowk}$ potential at different cutoffs. 
For the lower cutoffs the multipole values are converged at 10$\hbar\omega$, 
a larger value is necessary for higher cutoff. As expected, with the many-body correlations included the cutoff dependence gets weaker
and the monopoles converged in terms of $\hbar\omega$ are fairly
the same for studied cutoffs. With the only exception of the $\lambda\tau=20$ quadrupole term, which however, as
we show in the next paragraph, has a minor impact on the calculated properties of deformed nuclei.

Finally, in Tab. \ref{tab-conv} we show the convergence of 
calculated multipoles in the $sd$-shell in the perturbation expansion. Tha calculation is done for the $V_{lowk}$
cutoff $\Lambda=2.0$fm$^{-1}$.

\begin{table}
\begin{center}
\caption{Convergence of the eigenvalues of different multipole terms
in the order-by-order perturbative expansion. $V_{lowk}$ states for the first order interaction, 
CP means the second-order interaction with the core-polarization
contribution, 2nd-the second order interaction with all possible contributions (CP and 4p2h and ladder diagrams and folded diagrams),
$\hat Q (2)$ is the second order $\hat Q$-box with folded diagrams and $\hat Q(3)\equiv V_{eff}$ is the 3rd order $\hat Q$-box
interaction with folded diagrams.\label{tab-conv}}
\begin{tabular}{ccccccc}
\hline
\hline
Multipole & $V_{lowk}$ & CP & 2nd & $\hat Q(2)$ &3rd & $\hat Q(3)$ \\
\hline
$JT=10$   &  -4.93 & -4.48 & -8.23& -6.90& -10.05 &  -8.08 \\
$JT=01$   &  -4.41 & -6.31 & -7.41& -6.29& -7.16  &  -6.38 \\
\hline
$\lambda\tau=10$ & -1.53 & -2.11 & -1.93&-1.71 & -2.37 & -2.09 \\
$\lambda\tau=20$ & -2.86 & -3.43 & -4.58&-3.96 & -5.24 & -4.54 \\
$\lambda\tau=11$ &  2.75 &  3.15 &  3.80&3.35 &  4.68 &  4.09 \\
$\lambda\tau=30$ & -0.77 & -1.00 & -1.07&-0.94 & -1.02 & -1.10 \\
\hline
\hline
\end{tabular}
\end{center}
\end{table}

A large gain in the $JT=01$ channel between the first order interaction and that including CP
effects can be noticed from the Table, which improves the pairing properties obtained in the $V_{eff}$
substantially. Note, that nearly the same eigenvalue
of $JT=01$ is obtained for CP, $\hat Q(2)$ and $\hat Q(3)$. 
Otherwise, the order-by-order convergence is far from being satisfying for all the multipole terms.
A better convergence pattern is found in terms of $\hat Q$-boxes. As previously, 
the behaviour of $JT=10$ term appears the most unstable.

It is important to evoke that
the conclusions we obtain here on the convergence
of the multipole Hamiltonian follow exactly what has been discussed e.g. for pairing TBME derived from G-matrices
in Ref. \cite{Jensen95}, in $sd$ and $pf$ shells. Though there are conceptual differences
in the construction of the low$-k$ potentials and G-matrices, 
and in the resulting decoupling of high and low momenta in both methods,
these differences are not much apparent in the calculations in finite nuclei.
On the contrary, common problems are inherent to effective interactions derived from G-matrices and $V_{lowk}$
potentials, on both monopole \cite{Schwenk:2005me} and multipole levels.

\subsection{Multipoles and calculated observables}

To realize how the possible error in the multipoles of the Hamiltonian
influences the calculated observables, the following excercise was done: the monopoles
of $V_{lowk}$ and $V_{eff}$ (whose multipole eigenvalues are given in Tables \ref{tab-p},\ref{tab-sd})
were replaced by the monopoles of the corresponding $V_{emp}$. With such interactions, having exactly the same monopole
fields, we have recalculated the spectra of all nuclei which were used in the empirical fits of the Hamiltonian.
In the $sd$-shell we obtain the rms of 1.48MeV for $V_{lowk}$ and 1.18MeV $V_{eff}$, 
a very considerable deterioration from the rms=0.22MeV of the $V_{emp}$. Given that 
as an the intermediate step (see Sec. \ref{sec-fit}) we have 
fitted only monopoles of $V_{eff}$ and obtained the rms error of 0.83MeV, one might conclude that
the fine tuning of monopoles depends on the associated multipole and the correct 
multipole can be only defined within a given set of monopoles.

However, a stringent test for the pairing part of the Hamiltonian is provided
by the spectra of 2 particles (2 holes) systems, in which the $0^+-2^+$ splitting
should reflect the strength of the $J=0$ pairing interaction.
To some extent, the quadrupole content of the interaction can be also tested by properties of deformed
nuclei in a given model space.
Therefore, with the same interactions (i.e. having exactly the same monopoles) we have calculated the properties of 
$^{18}$O, $^{38}$Ar, $^{20}$Ne and $^{24}$Mg nuclei, which allows to visualize how the possible error 
in the multipole interaction influences the calculated observables.

The $2^+$ state of $^{18}$O, experimentally located at 1.98MeV, is of course accurately
reproduced by the empirical fit (1.97MeV) while the $V_{lowk}$ places it at 1.24MeV.
This anti-pairing behavior of $V_{lowk}$ is cutoff independent and also present in  
G-matrix interactions and taking into account
the core-polarization corrections is necessary to recover the correct 
multipole magnitude.  $V_{eff}$ does then much better in this context, 
leading to only a bit too high $2^+$ energy in $^{18}$O of 2.21MeV. In all three cases the 
wave function of the ground state contains $84\%$ of the $d_{5/2}^2$ configuration, which 
explains the sensivity of the calculated $2^+$ energy to the strength of the 
$d_{5/2}-d_{5/2}$ pairing interaction.
Further, we have calculated the $2^+$ energy of $^{38}$Ar. The ground state wave function in all cases
contains $\sim90\%$ of $d_{3/2}^{-2}$ component. The empirical interaction places the $2^+$
state at 1.86MeV (exp=2.16MeV), while the $V_{lowk}$ at a very low energy of 0.52MeV.
The $V_{eff}$ interaction, including CP effects, improves the agreement with experiment substantially, 
with the $2^+$ energy at 1.62MeV. One should note here, that although the eigenvalue
of the $JT=01$ is getting reduced from $V_{eff}$ to $V_{emp}$, this is not necessarily the trend of all
the pairing matrix elements; apparently the $d_{3/2}-d_{3/2}$ pairing is enhanced in the fit, contrary to the 
case of the $f_{7/2}-f_{7/2}$ one.

In Table \ref{tab-mgne} we show the quadrupole properties of $^{20}$Ne and $^{24}$Mg nuclei.
All the interactions give remarkably similar results, despite of differences
in their quadrupole content (see Tab. \ref{tab-sd}), a priori comparable to that
of the pairing part. This is a consequence of a more complex structure of deformed nuclei.
Since the wave functions in this case are spread over many components, the 20\% variation in the quadrupole
interaction does not lead to any considerable deviations, contrary to the cases of $^{18}$O and $^{38}$Ar, where
pairing dominates.

\begin{table}
\begin{center}
\caption{Quadrupole properties of $^{20}$Ne and $^{24}$Mg nuclei obtained with different interactions, see text for details. \label{tab-mgne}}
\begin{tabular}{llclll}
\hline
\hline
Nucleus & Observable && $V_{lowk}$ & $V_{eff}$ & $V_{emp}$ \\
\hline
$^{20}$Ne & $Q(2^+)$ (efm$^2$) && -15.7 & -15.9 & -15.8 \\
          & $B(E2; 2^+\rightarrow 0^+)$ (e$^2$fm$^4$)&& 59.8 & 59.7  & 59.3 \\
          & $B(E2; 4^+\rightarrow 2^+)$ (e$^2$fm$^4$)&& 72.9 & 70.9  & 71.8 \\
\hline
$^{24}$Mg &  $Q(2^+)$ (efm$^2$) &&-18.2 & -18.1 & -19.2\\
          &  $B(E2; 2^+\rightarrow 0^+)$(e$^2$fm$^4$)& &92.7  & 109.8 & 97.6  \\
          & $B(E2; 4^+\rightarrow 2^+)$(e$^2$fm$^4$) && 127.7 & 143.1  & 133.5 \\  
\hline
\hline
\end{tabular}
\end{center}
\end{table}

\section{Conclusions\label{sec-conc}}
We have performed the spin-tensor analysis 
of the monopole-free effective interactions
in the $p$ and $sd$ shells. We have shown that the leading (coherent) multipoles
of the realistic forces are dominated by the central interaction, 
supporting the equivalence of multipoles extracted from realistic TBME and schematic operator forms
proposed in Ref. \cite{Dufour:1995em}.
We have also shown that the renormalization persistency
of the tensor force, observed in the $J$-averaged matrix elements, is no longer present as
far as monopole-free matrix elements are concerned. However, this property of the tensor 
force is of a minor importance in the nuclear modeling as the contribution of the first-order tensor effects
to the coherent part of the multipole Hamiltonian is negligible. 
Further, we have compared the $V_{lowk}$ potentials based on the CD-Bonn
force to the phenomenological forces based on the same starting potential. 
The leading multipoles of the interactions are moderately modified
in the empirical fits, as required by the "realistic-compatibility" condition, nonetheless
their modification along with the monopole part is inevitable to obtain 
shell-model interactions of a satisfactory precision. On average, 
the leading multipoles of the empirical interaction appear weaker than in the $\hat Q$-box effective interactions
but stronger than in a realistic potential smoothed via the $V_{lowk}$ procedure
with a 2fm$^{-1}$ cutoff. 
The convergence properties of the multipoles of the effective interactions derived here
from the $V_{lowk}$ potentials are the same as those of the G-matrix based interactions from
Ref. \cite{Jensen95}, showing again their quantitative similarity invoked in Ref. \cite{Schwenk:2005me}.
It would be of a great interest to investigate how the inclusion of 3-body forces would
influence the couplings in the multipole Hamiltonian and wether it could explain the multipole modifications
required by the fit. Some light on this issue 
can be shed in the nearest future, once the 2-body effective interactions
with both, $T=1$ and $T=0$ effective 3-body components, are evaluated.

\section{Acknowledgements}
We are greatful to E. Caurier and F. Nowacki for 
certain numerical tools which were useful in the present study. 

%\bibliography{$HOME/BIB/kama}

\bibliographystyle{apsrev}

\end{document}